\documentclass[prb,showpacs,floatfix,%superscriptaddress,
groupedaddress,
longbibliography,
twocolumn]{revtex4-1}
\usepackage{graphicx}
\usepackage{epstopdf}
\usepackage{amssymb}
\usepackage{amsmath}
\usepackage{dsfont}
\usepackage{xspace}
\usepackage{color}
\usepackage{ulem}
\usepackage{xcolor}
\usepackage{hyperref}
\usepackage{subfigure}

\newcommand{\mean}[1]{\left\langle #1 \right\rangle}
\newcommand{\bra}[1]{\left\langle #1 \right\vert}
\newcommand{\ket}[1]{\left\vert #1 \right\rangle}

\newcommand{\carb}{$^{13}\mbox{C}$}
\newcommand{\NVm}{NV$^-$}

\newcommand{\mbf}[1]{\mathbf{#1}}

\begin{document}

\title{Dynamical quantum phase transition in diamond: applications in quantum metrology.}
\author{Francisco J. Gonz\'alez, Ariel Norambuena}
\affiliation{Centro de Investigaci\'on DAiTA Lab, Facultad de Estudios Interdisciplinarios, Universidad Mayor, Chile}
\author{Ra\'ul Coto}
\email{raul.coto@umayor.cl}
\affiliation{Centro de Investigaci\'on DAiTA Lab, Facultad de Estudios Interdisciplinarios, Universidad Mayor, Chile}
\date{\today}

\begin{abstract}
Nonequilibrium dynamics is a paramount scenario for studying quantum systems. The emergence of new features with no equilibrium counterpart, such as dynamical quantum phase transition (DQPT), has attracted wide attention. In this work, we depart from the well known Ising model and showcase an experimentally accessible configuration of a negatively charged Nitrogen-Vacancy center that interacts with nearby Carbon-13 nuclear spins. We provide new insights into this system in the context of DQPT. We show that nuclear spins undergo DQPT by appropriately choosing the relation between the transverse and longitudinal components of an external magnetic field. Furthermore, we can steer the DQPT via a time-dependent longitudinal magnetic field and apply this control to enhance the estimation of the coupling strength between the nuclear spins. Moreover, we propose a novel quenched dynamics that originates from the rotation of the central electron spin, which controls the DQPT relying on the anisotropy of the hyperfine coupling.
\end{abstract}

\maketitle

\section{Introduction.}

Quantum phase transitions (QPT) rank among the most striking behavior of matter, in which a quantum system experiences a sudden change of its properties~\cite{Sachdev11}. The variation of control parameters drives the system through a critical point where the free energy function becomes nonanalytic. In temperature-driven QPT, this point belongs to a critical temperature. However, QPT may happen even at zero temperature as quantum fluctuations drive the system's ground state. Furthermore, QPT can be susceptible to microscopic control parameters such as the atom-cavity detuning in Mott Insulator-Superfluid phase transitions~\cite{Toyoda,Greiner}. In general, equilibrium QPT are well understood and provide a suitable path for unravelling the properties of a system~\cite{Toyoda,Greiner,Hwang_2016}. In contrast, nonequilibrium QPT belong to an ongoing field that poses new challenges and opens new avenues to study QPT~\cite{Diehl10,heyl2018,Huber_2020,Tancara21}.

In recent years, nonequilibrium QPT have been studied in the context of physical quantities that become nonanalytic in time under quenched dynamics. This particular behavior has been termed as dynamical quantum phase transition (DQPT)~\cite{heyl2013}. Here, the time evolution resembles the effect of the driven parameter~\cite{heyl2013,heyl2018}. This idea has opened new horizons for theoretical studies about magnetization and entanglement~\cite{jurcevic2017}, parameter estimation~\cite{Invernizzi08,Sun2010}; as well as proof-of-principle experiments~\cite{jurcevic2017,Flaschner18,Zhang17}. Most of these studies have focused on the transverse-field Ising model~\cite{heyl2013,jurcevic2017,Gurarie}, where the quench occurs between the nearest-neighbor interaction ($\sigma_z^{i}\sigma_z^{i+1}$) and on-site interaction ($\sigma_x^{i}$). However, other configurations leading to experimentally accesible devices must be explored in order to gain more insights and control of this phenomena.

The negatively charged Nitrogen-Vacancy (\NVm) center in diamond is a promising platform for quantum technologies \cite{Atature2018, Awschalom2018}. The \NVm is an alternative to the well established platforms of superconducting qubits, trapped ions, and cold atoms. It has delivered important applications in quantum information processing~\cite{Wrachtrup06,Hensen2015}, quantum sensing\cite{Maze08aa,Balasubramanian_2008,Degen_2017,Coto2021} and quantum control \cite{Zhou17,Coto2017,Abobeih19,Gonzalez22}. Furthermore, it provides a testbed for different configurations of electron and nuclear spins~\cite{Waldherr2014,Taminiau2014,Abobeih19}.

In this work we show that the \NVm can be used to control surrounding Carbon-$13$ (\carb) nuclear spins to undergo DQPT. We extend the simulation of the well known Ising model to consider dipolar interactions between \carb, and anisotropic coupling to the \NVm. Moreover, we consider an off-axis magnetic field to steer the DQPT. In addition, dynamical steering is allowed through a time-varying field, revealing new insights into the nonequilibrium dynamics of color centers in diamond. In this direction, we show that this particular dynamics can be harnessed to deliver a quantum sensing protocol. Furthermore, we show that after freezing the dynamics (both the magnetization and the rate function reach a steady state), backstage dynamics of quantum correlations provides a time window for maximally entangled states between the \carb. %Finally, we account for the effect of transverse relaxation due to magnetic noise.

The paper is organized as follow. In section \ref{Sec_Spin} we introduce the system that is based on a \NVm interacting with nearby Carbon-13 nuclear spins. In section \ref{Sec_dqpt} we introduce DQPT, and we describe two different quenched dynamics, namely: quenched by external fields and quenched by a central spin. In section \ref{Sec_steering} we introduce a time-varying magnetic field to steer the DQPT. We apply this mechanism in quantum metrology to determine the coupling strength between two \carb. Furthermore, we prepare a steady state for the magnetization and show that quantum correlations build up to create a maximally entangled state. In section \ref{Sec_conclusions} we provide the final remarks of this work.

\section{Spins configuration.}\label{Sec_Spin}

\begin{figure}[t]
\centering
\includegraphics[width=230pt]{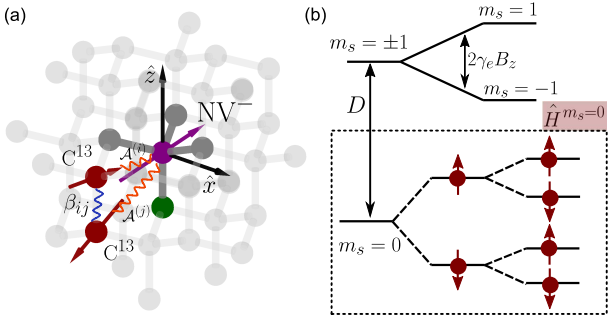}
\caption{(a) The \NVm interacts with two \carb~nuclear spins in the diamond lattice. $\mathcal{A}^{(i)}$ is the hyperfine coupling tensor between the \NVm and the $i$th nuclear spin. (b) Energy levels. The inset shows the splitting of the energy levels for the manifold $ m_{s} = 0$ when the \NVm interacts with two \carb~nuclear spins.}\label{Figure1}
\end{figure}

The negatively charged Nitrogen-Vacancy (\NVm) center in diamond is a point defect whose molecular structure is given by a substitutional nitrogen atom next to a vacancy in the crystal lattice. The \NVm center has a $C_{3v}$ symmetry and can be modeled as a two-electron hole system with spin $S=1$. From \textit{first-principles} calculations~\cite{Ivady14,Doherty14} based on dipole-dipole interaction of two electrons one obtains that the ground state of this defect presents a natural zero-field splitting $D/2\pi = 2.87 $ GHz between $m_s = 0$ and $m_s= \pm 1$ degenerated states~\cite{Jelezko04,Doherty12}. Hence, an external magnetic field along the N-V axis (symmetry axis) lifts the degeneracy between states $m_s= \pm 1$ because of the Zeeman effect. In addition, full control of the spin-triplet can be addressed by a microwave field~\cite{Jelezko04}. Nearby spin-$1/2$ ($I=1/2$) Carbon-13 (\carb) nuclear spins are hyperfine coupled to the \NVm. Moreover, each \carb~interacts with the nuclear spins bath via dipolar interaction, as shown in Fig.~\ref{Figure1}. The Hamiltonian of the system reads,
\begin{equation}\label{eq1}
\hat{H}=D\hat{S}_{z}^{2}+\gamma_{e}\mathbf{B}\cdot\mathbf{S}+\mathbf{S}\cdot\sum_{i=1}^{N}\mathcal{A}^{(i)}\cdot\mathbf{I}^{(i)}+ \hat{H}_{I},
\end{equation}
where $ \gamma_e/2\pi=2.8 $ MHz/G is the gyromagnetic ratio of the electron spin, $ \mathbf{B}=B_{x}\hat{x}+B_{z}\hat{z} $ corresponds to the external magnetic field. The $B_x$ component will be used for external control of the dynamics of the whole system. $\mathcal{A}^{(i)} $ is the hyperfine tensor, and $\hat{H} _{I} $ is the Hamiltonian of nuclear spins, that is given by,
\begin{equation}\label{eq2}
\hat{H}_{I}=\gamma_{n}\mathbf{B}\cdot\sum_{i=1}^{N}\mathbf{I}^{(i)}+\hat{H}_{n},
\end{equation}
where $ \gamma_n/2\pi=1.07$ kHz/G is the gyromagnetic ratio of the \carb~and $\mathbf{I}^{(i)}$ is the nuclear spin operator of the $i$th \carb. $H_{n}$ corresponds to the dipole interaction between the nuclear spins, that is written as,
\begin{equation}\label{eq3}
\hat{H}_{n}=\sum_{i<j}\frac{\mu_{0}\gamma_{n}^{2}}{4\pi r_{ij}^{3}}\left(\mathbf{I}^{(i)}\cdot\mathbf{I}^{(j)}-\frac{3(\mathbf{I}^{(i)}\cdot\mathbf{r}_{ij})(\mathbf{r}_{ij}\cdot\mathbf{I}^{(j)})}{r_{ij}^{2}}\right),
\end{equation}
with $ \mu_0 $ the vacuum permeability, $r_{ij}$ is the distance between the $i$th and $j$th nuclear spins. The vector $ \mathbf{r}_{ij} $ in spherical coordinates reads $\mathbf{r}_{ij}=(r_{ij}^x,r_{ij}^y,r_{ij}^z)=(r_{ij}\sin\theta_{ij}\cos\phi_{ij},r_{ij}\sin\theta_{ij}\sin\phi_{ij},r_{ij}\cos\theta_{ij})$. Hence, the Hamiltonian for the nuclear interaction is written as,
\begin{equation}\label{eq4}
\hat{H}_{n}=\sum_{i<j}\frac{\mu_{0}\gamma_{c}^{2}}{4\pi r_{ij}^{3}}[\hat{A}_{ij}+\hat{B}_{ij}+\hat{C}_{ij}+\hat{D}_{ij}+\hat{E}_{ij}+\hat{F}_{ij}],
\end{equation}
with
\begin{align}\label{eq5}
\hat{A}_{ij}& = \hat{I}_{z}^{(i)}\hat{I}_{z}^{(j)}(1-3\cos^{2}\theta_{ij}),\nonumber\\
\hat{B}_{ij}& = -\frac{1}{4}[\hat{I}_{+}^{(i)}\hat{I}_{-}^{(j)}+\hat{I}_{-}^{(i)}\hat{I}_{+}^{(j)}](1-3\cos^{2}\theta_{ij}),\nonumber\\
\hat{C}_{ij}& = -\frac{3}{2} [\hat{I}_{+}^{(i)}\hat{I}_{z}^{(j)}+\hat{I}_{z}^{(i)}\hat{I}_{+}^{(j)}]\sin\theta_{ij}\cos\theta_{ij} e^{-i\phi_{ij}},\nonumber\\
\hat{D}_{ij}&=-\frac{3}{2} [\hat{I}_{-}^{(i)}\hat{I}_{z}^{(j)}+\hat{I}_{z}^{(i)}\hat{I}_{-}^{(j)}]\sin\theta_{ij}\cos\theta_{ij} e^{i\phi_{ij}},\nonumber\\
\hat{E}_{ij}& = -\frac{3}{4}\hat{I}_{+}^{(i)}\hat{I}_{+}^{(j)}\sin^{2}\theta_{ij} e^{-2i\phi_{ij}},\nonumber\\
\hat{F}_{ij}&= -\frac{3}{4}\hat{I}_{-}^{(i)}\hat{I}_{-}^{(j)}\sin^{2}\theta_{ij} e^{2i\phi_{ij}},
\end{align}

where $\theta_{ij}$ is the angle between $ \mathbf{r}_{ij}$ and the $ B_{z} $ component of the magnetic field, while $ \phi_{ij} $ is the azimuth angle with respect to the $ \hat{x}$ axis. For large magnetic fields the terms $ \hat{C}_{ij} $, $ \hat{D}_{ij} $, $ \hat{E}_{ij} $, $ \hat{F}_{ij} $ can be neglected under the so-called secular approximation. Then, the Hamiltonian~\eqref{eq4} reads,
\begin{equation}\label{eq6}
\hat{H}_{n}=\sum_{i<j}\frac{\beta_{ij}}{4}\left[(\hat{I}_{+}^{(i)}\hat{I}_{-}^{(j)}+\hat{I}_{-}^{(i)}\hat{I}_{+}^{(j)})-4\hat{I}_{z}^{(i)}\hat{I}_{z}^{(j)}\right],
\end{equation}
where $\beta_{ij}=-\frac{\mu_{0}\gamma_{c}^{2}}{4\pi r_{ij}^{3}}(1-3\cos^{2}\theta_{ij})$.

Considering the zero-field splitting to be larger than the perpendicular magnetic field and the hyperfine coupling, i.e. $D \gg \gamma_e B_x$ and $D \gg \mathcal{A}^{(i)}$ , one can perform the secular approximation, that neglects $\hat{S}_x$ and $\hat{S}_y$ contributions in the second and third terms in the Hamiltonian \eqref{eq1}. We numerically confirm this approximation and consider transverse relaxation in Appendix~\ref{Appendix_Secular}. Hence, the Hamiltonian for the tripartite system can be written conditioned to the electron spin manifold, such that,
\begin{align}\label{eq7}
\hat{H}^{m_s}&=(m_s^2 D+m_s\gamma_eB_z)+\gamma_n\mathbf{B}\cdot\sum_{i=1}^{N}\mbf{I}^{(i)}\nonumber\\
&\quad+m_{s}\sum_{i=1}^{N}\sum_{\alpha=x,y,z}\mathcal{A}_{z\alpha}^{(i)}\hat{I}_{\alpha}^{(i)}\quad+\hat{H}_{n}.
\end{align}

\section{Dynamical quantum phase transition.}\label{Sec_dqpt}

Nonequilibrium phase transitions give rise to different dynamics that in several cases have no equilibrium counterpart~\cite{Vajna,heyl2018}. The nonequilibrium dynamics originates from different scenarios like Floquet engineering~\cite{Bastidas2012,johansen2021,Zamani20}, reservoir coupling~\cite{Minganti,Fink2018}, quenched parameters~\cite{heyl2013,Flaschner18}, among others. In this work, we focus on a quenched dynamics that hereafter we will refer to as dynamical quantum phase transition (DQPT) \cite{heyl2013}. For this goal, we consider a Hamiltonian of the form, $\hat{H}=\hat{H}_{0}+\hat{H}_{1}$, where $ \hat{H}_{0}$ has two degenerated eigenstates, namely $\ket{\Downarrow}$ and $\ket{\Uparrow}$. The system is initially prepared in one of the $\hat{H}_0$ eigenstates, say $\ket{\Downarrow}$, and suddenly $\hat{H}_{1}$ is turned on.

It has been shown that the Loschmidt amplitude plays a pivotal role in DQPT \cite{heyl2013,heyl2018,Vajna,Canovi}, resembling the canonical partition function in equilibrium phase transition. The Loschmidt amplitude reads,
\begin{equation}\label{eq12}
\mathcal{G}(t)=\bra{\varPsi(0)}e^{-i\hat{H}t}\ket{\varPsi(0)},
\end{equation}

and it gives the projection of the time evolved state with the initial state, that we chose to be a ground state. It is also convenient to introduce the Loschmidt echo, that it is interpreted as the return probability to the ground state manifold (two-fold degenerated) $\mathcal{L}(t)=P(t)=P_{\Downarrow}(t)+P_{\Uparrow}(t)$, with $P_{i}=|\langle i | e^{-i\hat{H}t}\ket{\varPsi(0)}|^2$, and $i=\{\Downarrow, \Uparrow\}$~\cite{jurcevic2017,heyl2018,andraschko2014}. For simplicity, we restrict our analysis to the case where $\mathcal{L}(t)$ exhibits an exponential dependence upon the number of degrees of freedom $N$. Therefore, we can introduce a rate function as,
\begin{equation}\label{eq11}
\Lambda(t)=-\lim_{N\rightarrow \infty}\frac{1}{N}\log[P(t)].
\end{equation}

Then, in analogy with the free energy potential, the nonanaliticity of $\Lambda(t)$ at the critical times $t_c$ probes the DQPT. However, the above expression for $\Lambda(t)$ considers the thermodynamic limit ($N\rightarrow \infty$), which is a drawback for experiments and quantum simulations. Instead, it is worth considering the dominant contribution of the probability, that yields our main tool to predict DQPT for small systems~\cite{jurcevic2017,zunkovic2018,heyl2014},
\begin{equation}\label{eq13}
\lambda(t)=\min_{\eta\in\{\Downarrow,\Uparrow\}}\left(-\frac{1}{N}\log[P_{\eta}(t)]\right).
\end{equation}

We remark that $\lambda(t)$ coincides with $\Lambda(t)$ for large $N$~\cite{heyl2018}, and it shows the nonanaliticity when crossing the region $P_{\Uparrow}=P_{\Downarrow}$. Here, the dynamics restores the symmetry in the ground state probability $P(t)$ initially broken by the state preparation.

Order parameters are crucial to witness quantum phase transitions. In general, one seeks observables that highlight differences between the phases, and exhibit a sudden change when crossing the critical points. Nonequilibrium dynamics demands dynamical order parameters to account for critical times~\cite{jurcevic2017,heyl2018,heyl2017}. Firstly, we will focus on the magnetization as the dynamical order parameter, $\mean{M_{z}} =(1/N) \sum_{i = 1}^{N} \mean{\hat{I}_{z}^{ (i)}} $, and later on we will discuss the case of quantum correlations, see Appendix~\ref{Appendix_Entanglement}. The former, has signaled DQPT by vanishing when the system restores the symmetry~\cite{heyl2018}.

\subsection{Quenched dynamics by longitudinal and transverse magnetic fields.}

First, we focus on the $ m_ {s} = 0 $ manifold of the \NVm~electron spin in Eq.~\eqref{eq7}. Furthermore, considering $\theta_{ij} = 0$ we obtain the following Hamiltonian,
\begin{align}\label{eq8}
\hat{\mathcal{H}}&=\frac{\beta_{12}}{2}\left((\hat{I}_{+}^{(1)}\hat{I}_{-}^{(2)}+\hat{I}_{-}^{(1)}\hat{I}_{+}^{(2)})-4\hat{I}_{z}^{(1)}\hat{I}_{z}^{(2)}\right)\nonumber\\
&+\gamma_nB_{z}(\hat{I}_{z}^{(1)}+\hat{I}_{z}^{(2)})+\gamma_nB_{x}(\hat{I}_{x}^{(1)}+\hat{I}_{x}^{(2)}),
\end{align}

where $\beta_{12}=\mu_{0}\gamma_{c}^{2}/(4\pi r_{12}^{3})$ and $\hat{I}_{\pm}^{(i)}=\hat{I}_{x}^{(i)}\pm i\hat{I}_{y}^{(i)}$. We identify $\hat{H}_0$ and $\hat{H}_1$ with the first and second line in Eq.~\eqref{eq8}, respectively. Hence, in the absence of magnetic field ($t=0$), the dynamics is governed by $ \hat{H}_{0}$ which has two degenerate eigenstates, namely $\ket{0\downarrow\downarrow}=\ket{\Downarrow}$ and $\ket{0\uparrow\uparrow}=\ket{\Uparrow}$.

\begin{figure}[t]
\centering
\includegraphics[width=250pt]{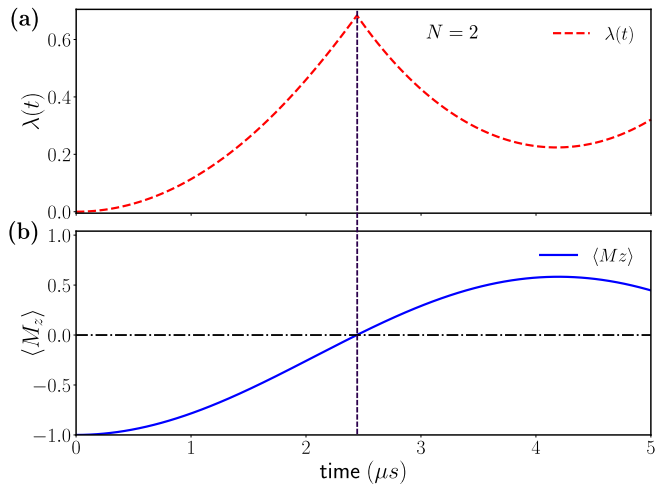}
\caption{Nonanalytical points in the rate function $ \lambda (t) $ (a) coincides with vanishing magnetization $ \mean{M_{z}} $ (b) at critical times. For the simulation we consider two nuclear spins $ (N = 2) $ and magnetic fields $ B_{x} = 100 $ G and $ B_{z} = 50 $ G.}\label{Figure2}
\end{figure}

In Fig. \ref{Figure2}-(a) we show that \carb~nuclear spins surrounding the \NVm center undergo a DQPT. The DQPT is witnessed through nonanalyticities in the rate function $\lambda(t)$ at the critical time $t_{c_1}=2.4$ $\mu$s. Furthermore, in Fig. \ref{Figure2}-(b) we show the evolution of the magnetization (dynamical order parameter) that, as mentioned above, vanishes at the critical time $t_{c_1}$. For completeness, we show in Appendix~\ref{Appendix_Scalability} that the critical time holds when increasing the number of nuclear spins.
 
\begin{figure*}[htb]
\centering
\includegraphics[scale=0.3]{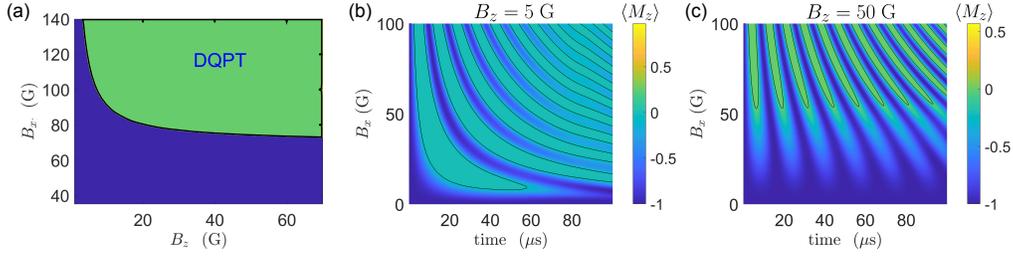}
\caption{Phase diagrams for the dynamical quantum phase transition. In panel (a) we vary magnetic fields $B_x$ and $B_z$ to find the regions where dynamical quantum phase transition exists (does not exist), which is illustrated by the yellow (blue) colored area. The magnetization as a function of time and the magnetic field $B_x$ is calculated for $B_z = 5$G (b) and $B_z = 50$G (c). The solid black lines shows the regions where $\langle M_z \rangle =0$.}
\label{Figure3}
\end{figure*}

In last years, the transverse Ising model has been a playground for the study of dynamical quantum phase transitions~\cite{heyl2013,jurcevic2017,Gurarie}. However, color centers in diamond offer an ubiquitous solid-state platform for extending this model. The central spin given by the electron spin of the \NVm serves as a control qubit upon the nuclear spins~\cite{Cooper2020,Gonzalez22}. Moreover, dipolar coupling between nuclear spins raises a complex dynamics that can be further controlled with a bias off-axis magnetic field. We numerically show here that the relation between the transverse ($B_z$) and longitudinal ($B_x$) magnetic fields define the region where DQPT takes place, as shown in Figure~\ref{Figure3}-(a). The bottom region, where the DQPT fails, is characterized by a negative average of the magnetization. For illustration, in Figure~\ref{Figure3}-(b),(c), we show the magnetization as a function of $B_x$ and time for fixed amplitudes of $B_z$. When moving ahead in time, one can observe multiples DQPT given by the zero magnetization threshold. Nevertheless, by increasing the transverse field up to $B_z=50$ G, we observe that there is a broader region for the longitudinal field where no DQPT appears (Fig.~\ref{Figure3}-(c)).

\subsection{Quenched dynamics by a central spin.}\label{Sec_central}

A new depart from quenched parameters can be worked out with a sudden change of the electron spin manifold. To our best knowledge, this provide a new route towards studying DQPT. The \NVm~acting as a central spin conditions the \carb~nuclear spins Hamiltonian, beyond the transverse Ising model. In contrast to the previous section, here the transverse and longitudinal magnetic fields are replaced by the isotropic and anisotropic hyperfine couplings. In the absence of the magnetic fields, we end up with a degenerated two-level \NVm electron spin ($\vert m_s=0\rangle=\vert0\rangle$ and $\vert m_s=\pm\rangle=\vert1\rangle$). A nucleus-independent rotation (hard $\pi$-pulse) on the electron spin can be achieved with a Rabi frequency $\sim 8$ MHz~\cite{Blok14}, which allows us to consider the rotation to be instantaneous, as compared with the dipolar interaction $\hat{H}_0$ (first line in Eq.~\eqref{eq8}). We begin by preparing the system in one of the $\hat{H}_0$ eigenstates, $\vert 0\downarrow\downarrow\rangle$. Then, the quenched dynamics originates from the instantaneous rotation of the electron spin to state $\vert1\rangle$, that transforms the Hamiltonian to $\hat{H}=\hat{H}_0+\hat{H}_1$, with $\hat{H}_1$ given by
\begin{align}
\hat{H}_1&=D\hat{S}_{z}^{2}+\hat{S}_{z}\mathcal{A}_{zz}^{(1)}\hat{I}_{z}^{(1)}+\hat{S}_{z}\mathcal{A}_{zz}^{(2)}\hat{I}_{z}^{(2)}\nonumber\\
&\quad+\frac{1}{2}\hat{S}_{z}\mathcal{A}_{ani}^{(1)}(\hat{I}_{+}^{(1)}e^{-i\phi_{1}}+\hat{I}_{-}^{(1)}e^{i\phi_{1}})\nonumber\\
&\quad+\frac{1}{2}\hat{S}_{z}\mathcal{A}_{ani}^{(2)}(\hat{I}_{+}^{(2)}e^{-i\phi_{2}}+\hat{I}_{-}^{(2)}e^{i\phi_{2}}),
\end{align}

where $\mathcal{A}_{ani}^{(i)}=({\mathcal{A}_{zx}^{(i)}}^2 + {\mathcal{A}_{zy}^{(i)}}^2)^{1/2}$ and $\tan\phi_i=\mathcal{A}_{zy}^{(i)}/\mathcal{A}_{zx}^{(i)}$.

\begin{figure}[h]
\centering
\includegraphics[width=240pt]{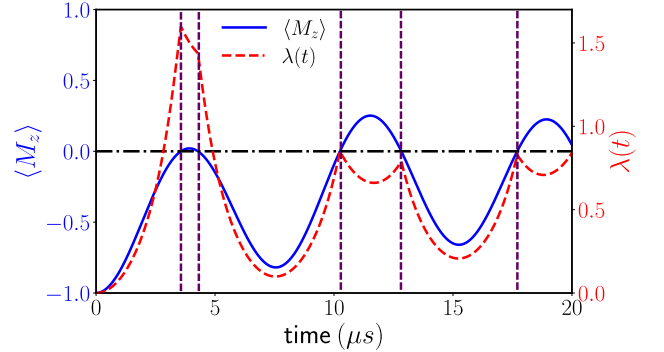}
\caption{The quenched dynamics induced by the central spin (\NVm) leads to DQPT, as witnessed by the rate function (dashed red line) and magnetization (solid blue line). The dotted vertical lines correspond to the times where the rate function is nonanalytical and coincides with a vanishing magnetization.}\label{Figure4}
\end{figure}

Next, we study DQPT as the probability to return to the initial state in the nuclear spins manifold ($\ket{\downarrow\downarrow}$) after the time evolution under the total Hamiltonian $\hat{H}$ for the electron spin manifold $m_{s}=\pm1$. In Fig.~\ref{Figure4} we show the rate function $\lambda(t)$ and the magnetization $\langle M_z\rangle$ for the following set of \carb~nuclear spins, $\mathcal{A}_{zz}^{(1)}=-27$ kHz, $\mathcal{A}_{zz}^{(2)}=-28$ kHz, $\mathcal{A}_{ani}^{(1)}=128$ kHz and $\mathcal{A}_{ani}^{(2)}=19$ kHz~\cite{Dreau14}.

Our simulations reveal a DQPT for the novel quenching we are considering here. It is important to notice that the DQPT appears when the anisotropic hyperfine coupling ($\mathcal{A}_{ani}^{(i)}$) is greater than the isotropic one ($\mathcal{A}_{zz}^{(i)}$). We support this statement with numerical simulations with the parameters sets reported in Refs.~\cite{Dreau14,Reiserer2016}. In contrast, the rate function does not present nonanalytical points and the magnetization always remains negative when the anisotropic component is weaker than the isotropic one, as observed for $\mathcal{A}_{zz}^{(1)}=2.281$ MHz, $\mathcal{A}_{zz}^{(2)}=1.884$ MHz, $\mathcal{A}_{ani}^{(1)}=0.240$ MHz and $\mathcal{A}_{ani}^{(2)}=0.208$ MHz~\cite{Nizovtsev2014}.

\begin{figure*}[ht]
\centering
\includegraphics[width=400pt]{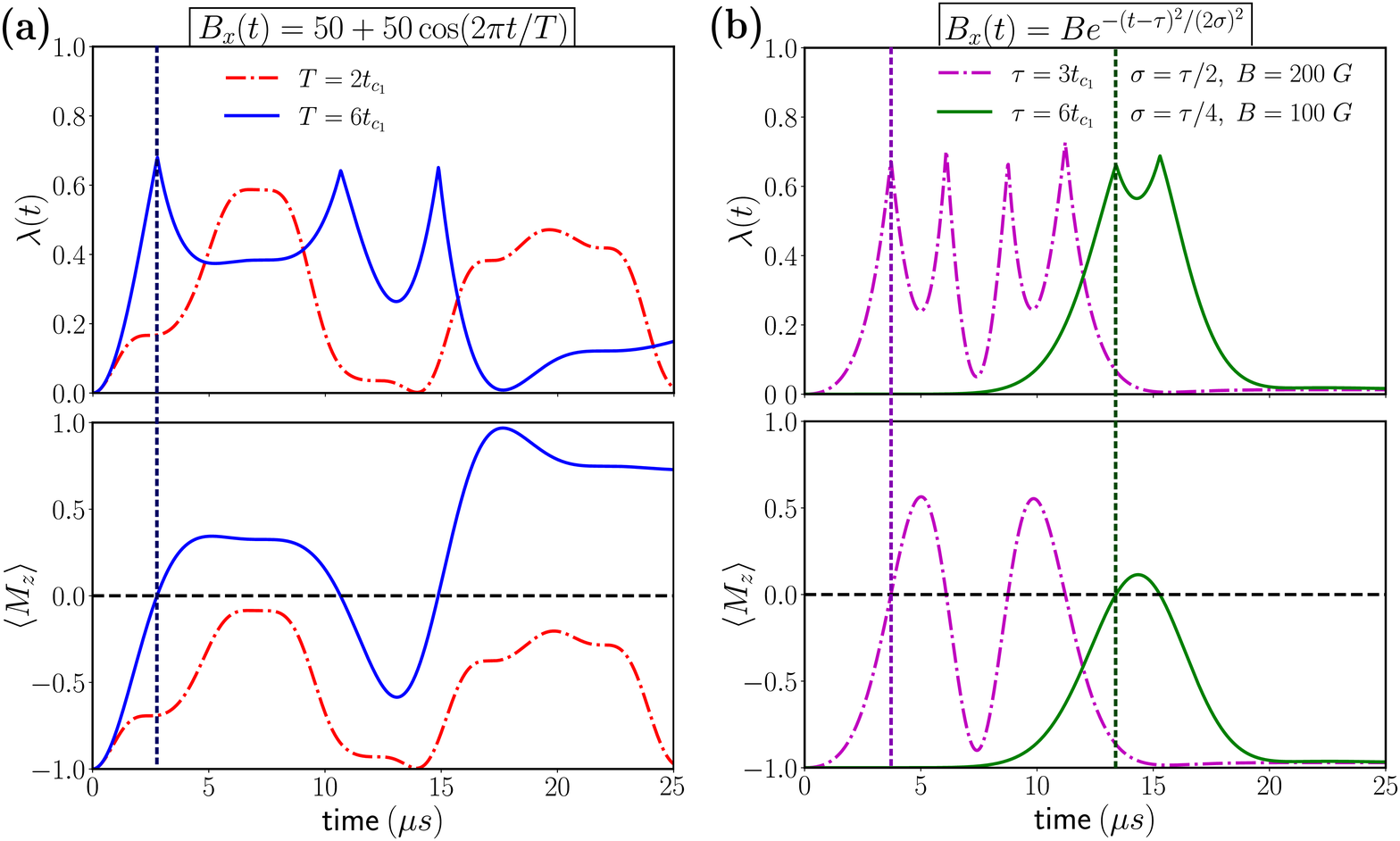}
\caption{Controlling dynamical quantum phase transition via time-varying magnetic fields. In the top (bottom) panels we show the rate function $\lambda(t)$ (magnetization $\langle M_z \rangle$). (a) With an oscillatory field $B_x(t)$ we are able to steer the DQPT by controlling the period. Here, we reach a maximum field $B_x(t)$ that amounts to $100$ G. $t_{c_1}$ is the critical time for the rate function when the magnetic field is fixed to $B_x= 100$ G, see Fig.~\ref{Figure2}. (b) With a Gaussian field we have more control on the DQPT, where we can change the critical time and the number of critical points. For all the simulations we considered a closed dynamics under the secular approximation with the initial state $ |0\downarrow \downarrow \rangle$, and $B_z = 50$ G.}\label{Figure5}
\end{figure*}

\section{Dynamical steering via time-varying magnetic field.}\label{Sec_steering}

For concreteness, hereafter we focus on the quenched dynamics triggered by the magnetic fields. The competition between the longitudinal and transverse magnetic fields allows us to dynamically steer the DQPT. Let us begin by considering the Hamiltonian $\hat{H}$ in Eq.~\eqref{eq8} with a time-dependent magnetic field, $B_x(t)$, and we calculate the probabilities according to this time-dependent Hamiltonian. 

In what follows, we study the effect of two different fields,
\begin{eqnarray}
B_x(t) &=& B_{x0} + A \cos\left({2 \pi t \over T}\right), \hspace{0.4 cm} \mbox{oscillating field}, \label{Bx_osc} \\
B_x(t) &=& B e^{-(t-\tau)^2/(2\sigma^2)}, \hspace{1.3 cm} \mbox{Gaussian field}. \label{Bx_gauss}
\end{eqnarray}

First, the oscillating field represents a sinusoidal signal that oscillates around the value $B_{x0}=50$ G with period $T$ and amplitude $A=50$ G. Second, the Gaussian field describes a localized pulse around the time $\tau$ with a characteristic width $\sigma$ and amplitude $B$. In Fig.~\ref{Figure5} we show the behavior of the rate function $\lambda(t)$ (top panel) and the magnetization $\langle M_z \rangle$ (bottom panel). From the oscillating field, Fig.~\ref{Figure5}-(a), we note that depending on the period, the system may undergo a DQPT. To understand this, we remark that our setting for the fields is similar to the one in Fig.~\ref{Figure2}, where the first DQPT takes place at the critical time $t_{c_1}\approx 2.4$ $\mu$s for a constant field $B_x=100$ G. Considering the oscillating field for $T=2t_{c_1}$, we realize that the accumulated action of the field $\int_{0}^{t_{c_1}}dt\, B_x(t)$ is smaller than that for the constant field, and hence there is no DQPT. For the case $T=6t_{c_1}$, the action of the field occurs for a longer time, which yields a DQPT. The main conclusion here is that the accumulated action of the magnetic field can be used to control whether the DQPT occurs by sweeping this quantity near a threshold region.

The above result enables us to define the control in terms of an effective area. Therefore, the Gaussian field is more suitable since it steers the DQPT by allowing us to set the time where the DQPT occurs, and also the number picks (crossings) in $\lambda$ ($\langle M_z \rangle$), see Fig.~\ref{Figure5}-(b). To summarize, we can steer the DQPT in a nonequilibrium dynamics by controlling a time-dependent magnetic field.

\subsection{Applications in Quantum Metrology.}

Spin-spin interaction is a central topic in quantum physics, and the precise knowledge of the strength of this interaction is a key aspect. However, there are several approaches for estimating the coupling strength, each one with pros and cons depending on the system, noises, measurement apparatus, etc. In particular, the estimation of the coupling strength between two \carb~nuclear spins in diamond has attracted attention. For instance, in the seminal work in Ref.~\cite{Jiang09}, the authors considered a simple sequence comprising the initialization of the tripartite system (\NVm and two \carb), followed by time evolution and subsequent measurement on one of the nuclear spins. A more elaborated scheme based on weak measurement has been proposed to estimate the hyperfine coupling between the \NVm and a \carb~\cite{Shikano2011}, which could be also extended to determine the interaction between two \carb. In this work, we provide an alternative viewpoint for this task, which involves a proof-of-principle demonstration of the role of nonequilibrium dynamics for quantum metrology. Previous works dealing with parameter estimation around critical points have focused on the Ising model and delivered opposite outcomes for slightly different purposes. Here, we contribute to the ongoing debate by showing that DQPT provides an advantage for quantum metrology.

In what follows, we use the Fisher Information (FI) to quantify the amount of information that can be retrieved from the dipolar coupling strength ($\beta_{12}$) between the two \carb~for a particular measurement scheme. FI of an unknown parameter $x$ is defined as
\begin{equation}\label{FI}
\mbox{FI}(x)=\sum_{i}\frac{1}{P_{i}(x)}\left(\frac{dP_{i}(x) }{d x}\right)^{2},
%\mbox{FI}(x)=\sum_{i}\frac{1}{P_{i}(d_{12})}\left(\frac{d }{d d_{12}}P_{i}(d_{12})\right)^{2},
\end{equation}
where $P_{i}(x)$ is the probability of the measurement outcome $i$ and the sum is over all the outcomes. Our measurement strategy involves measurements on one of the nuclear spins, which provides two possible outcomes, the probability of being in spin up $(P _ {\ket{\uparrow}} (\beta_{12}))$ or spin down ($P_{\ket{\downarrow}} (\beta_ {12}) = 1- P_ {\ket{\uparrow}}(\beta_{12}))$. Hence, the FI reduces to
\begin{equation}\label{FI2}
\mbox{FI}(\beta_{12})=\frac{1}{P_{\ket{\uparrow}}(\beta_{12})(1-P_{\ket{\uparrow}}(\beta_{12}))}\left[\frac{d P_{\ket{\uparrow}}(\beta_{12})}{d \beta_{12}}\right]^{2}.
\end{equation}

\begin{figure}[h]
\centering
\includegraphics[width=230pt]{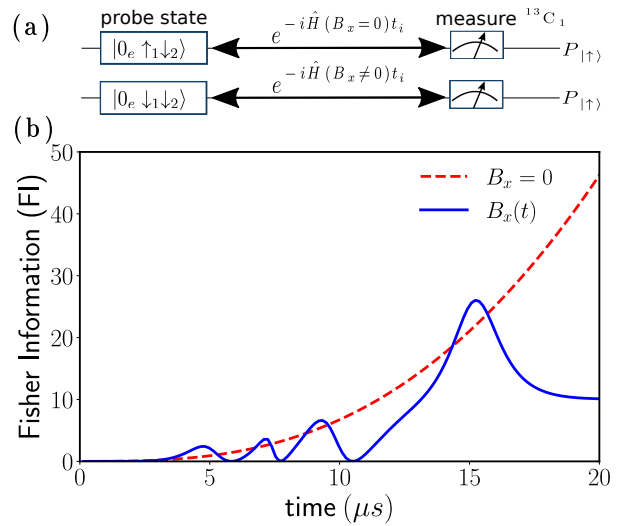}
\caption{(a) Two different measurement strategies for the Fisher Information, the second involves DQPT. (b) DQPT driven by a Gaussian longitudinal magnetic field improves the Fisher Information corresponding to the dipolar coupling $\beta_{12}$, as compared to the case without DQPT at $B_x=0$. We used $ B_ {x}(t) =B \exp\{-(t-\tau)^2/(2\sigma^2)\} $, with $B=200$ G, $\tau=3t_{c_{1}}$ and $\sigma = \tau/2$. $B_{z}=50$ G is fixed for both cases.}\label{Figure6}
\end{figure}

The measurement strategy considers the initialization of the system in a probe state, the time evolution under the Hamiltonian $\hat{H}$ in Eq.~\eqref{eq8} for a certain interrogation time ($t_i$), followed by a measurement on the nuclear spin (see Fig. \ref{Figure6}-(a)). We analyze two particular cases for the longitudinal component of the magnetic field and probe state $ i) $ $ B_{x} = 0$ with $\ket{\psi(0)}=\ket{0\uparrow\downarrow}$; and $ii)$ $B_x(t) = B e^{-(t-\tau)^2/(2\sigma^2)}$ with $\ket{\psi(0)}=\ket{0\downarrow\downarrow}$. The transverse component remains constant for both cases ($B_{z}=50$ G).

In Fig.~\ref{Figure6}-(b) we show the evolution of the FI. We observe that in the absence of a longitudinal magnetic field $B_{x}=0$ (resembling the protocol in Ref.~\onlinecite{Jiang09}), the FI follows a quadratic evolution, $\mbox{FI}=t^{2}$. Details of the calculation are given in Appendix~\ref{Appendix_Fisher}. In the presence of the Gaussian field (in the DQPT regime shown in Fig.~\ref{Figure5}-(b)), the FI shows oscillations that outperform the previous result. We remark that the nonequilibrium dynamics is crucial for this enhancement since a constant magnetic field $B_x$ delivers no improvement. Hence, we demonstrate that improved metrology can be attained in color centers in diamond by driving the system around critical points.

\subsection{Quantum correlations in stationary magnetization.}

Another important case of nonequilibrium dynamics appears when studying the steady state of the system in terms of the order parameter, i.e. the asymptotic behavior of the magnetization. For instance, this problem has been addressed in the transverse Ising model with long range interactions \cite{Heyl2018bb}, where the authors found a connection between DQPT and this nonequilibrium criticality (that is another kind of DQPT). In this section, we show that even when the magnetization reaches a steady state, quantum correlations (nondiagonal elements of the density matrix) oscillate between the maximum and minimum values of Concurrence~\cite{Wootters1998}.

\begin{figure}[ht]
\centering
\includegraphics[width=220pt]{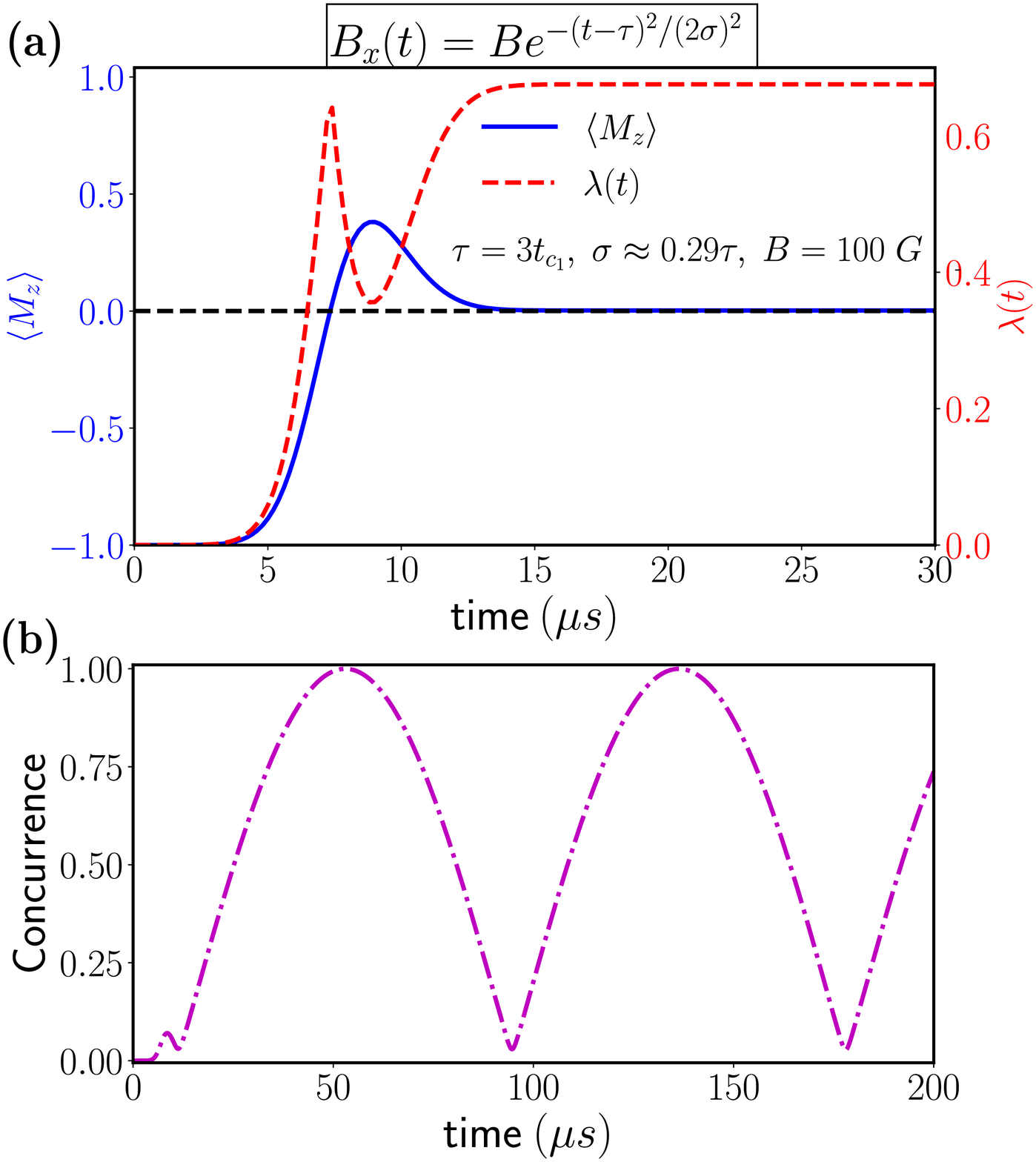}
\caption{(a) Evolution of the rate function $ \lambda(t)$ (dashed red line) and magnetization (solid blue line) for a Gaussian field. The magnetization reaches a steady state that is adjusted around $\langle M_z\rangle =0$. (b) The Concurrence shows oscillations even when the magnetization has been frozen, and amounts to one, indicating the creation of a maximally entangled state.}\label{Figure7}
\end{figure}

In Fig.~\ref{Figure7}-(a) we show the time evolution of the magnetization and the rate function. Fig.~\ref{Figure7}-(b) shows the evolution of the Concurrence in the same interval, where a periodic behavior can be observed while the magnetization is in a stationary state. From Fig.~\ref{Figure7}-(b) we retrieved the disentangled and the maximally entangled states to be of the form
\begin{equation}
\ket{\psi}=r\ket{0}\otimes\left(e^{i\varphi_1}\ket{\uparrow\uparrow}+e^{i\varphi_2}(\ket{\uparrow\downarrow}+\ket{\downarrow\uparrow})+e^{i\varphi_3}\ket{\downarrow\downarrow}\right),
\end{equation}

with $r \approx1/2$. On the one hand, the maximally entangled state ($\mathcal{C}\approx 1$~\cite{Wootters1998}) is given by $\lbrace \varphi_1=0.31,\varphi_2=1.16,\varphi_3=-1.13\rbrace$. On the other hand, the disentangled state ($\mathcal{C}\approx0$) is $\lbrace \varphi_1=-0.59,\varphi_2=0.11,\varphi_3=0.82\rbrace$. The definition and other calculations with the Concurrence ($\mathcal{C}$) are given in Appendix~\ref{Appendix_Entanglement}.

\section{Conclusions}\label{Sec_conclusions}

In this work we proposed a negatively charged Nitrogen-Vacancy (\NVm) center and nearby Carbon-$13$ (\carb)~nuclear spins as a testbed for studying dynamical quantum phase transition (DQPT). We found two different quenched dynamics that enforce DQPT on the nuclear spins. First, we observed that nuclear spins undergo DQPT by appropriately choosing the relation between the transverse and longitudinal components of an external magnetic field. Moreover, one can steer the DQPT via a time-dependent longitudinal magnetic field. In addition, this dynamical steering can be harnessed to enhance the Fisher Information concerning the estimation of the coupling strength between two \carb~nuclear spins. Second, by rapidly rotating the \NVm electron spin (that is a central spin), the \carb~nuclear spins undergo a DQPT depending on the relation between the anisotropic and isotropic coupling of the hyperfine interaction. We also studied the role of bipartite and tripartite entanglement during the critical points where DQPT takes place, and in the steady state of the magnetization. Overall, we believe that \NVm and surrounding nuclear spins provide a prototype for studying nonequilibrium dynamics, and in particular, DQPT.

\section{acknowledgments} FJG acknowledges support from Universidad Mayor through the Doctoral fellowship. A.N. acknowledges financial support from Universidad Mayor through the Postdoctoral Fellowship. RC acknowledges support from Fondecyt Iniciaci\'on No. 11180143.

\appendix

\section{Secular approximation and relaxation.}\label{Appendix_Secular}

Along the manuscript, we considered a lossless scenario and the secular approximation. The latter allows us to simplify the analytical calculations by restricting the Hamiltonian to be conditioned to the \NVm electron spin manifold. This approximation breaks down when $D \lesssim \gamma_e B_x$, and hence the Zeeman terms $\hat{S}_x$ and $\hat{S}_y$ must be considered. Furthermore, we shall consider magnetic noise on the \NVm and \carb~nuclear spins. When the electron spin is in the $m_s=0$ manifold, we considered it isolated from magnetic noise. However, when the electron spin occupies states $m_s=\pm1$ (see Section~\ref{Sec_central}) a transverse relaxation process must be taken into account~\cite{Coto2021,Gonzalez22}. Nevertheless, the \NVm electron spin coherence time typically ranges from $4$ to $10$ $\mu$s~\cite{Maze08aa}, which provides enough time to observe the DQPT in Fig.~\ref{Figure4}. To support our numerical calculations, we include the full Hamiltonian (without secular approximations) and transverse relaxation over the nuclear spins with a coherence time $T_{2n}^\star=0.5$ ms~\cite{Gonzalez22}. We find that for magnetic fields below $B_x=500$ G, our simplified model reproduce very well the magnetization up to $70$ $\mu$s. The rate function $\lambda(t)$, which is less important for the physical validation of the model, behaves well up to $20$ $\mu$s. The reason behind this is the logarithmic function in its definition that increases the mismatch.

\section{Entanglement as order parameter.}\label{Appendix_Entanglement}

In the past, it has been shown that critical points corresponding to DQPT yield increased quantum correlations~\cite{jurcevic2017}. Here, we contribute to this analysis by showing the same behavior for the entanglement, but also by shedding light on the multipartite entanglement. First, we quantify the entanglement by the Concurrence~\cite{Wootters1998}, that is defined as
\begin{equation}\label{Conc}
\mathcal{C}(\rho_{12})=\max(\lambda_{1}-\lambda_{2}-\lambda_{3}-\lambda_{4},0),
\end{equation}
where the $ \lambda_{i} $ are the square roots of the eigenvalues, in decreasing order, of the matrix $\mathbf{R}=\hat{\rho}_{12}(\hat{I}_{y}^{(1)}\otimes\hat{I}_{y}^{(2)})\hat{\rho}_{12}^{\ast}(\hat{I}_{y}^{(1)}\otimes\hat{I}_{y}^{(2)})$. $\hat{\rho}_{12}^{\ast}$ is the complex conjugate of the density operator of the bipartite system corresponding to the two \carb~nuclear spins.

\begin{figure}[h]
\centering
\includegraphics[width=250pt]{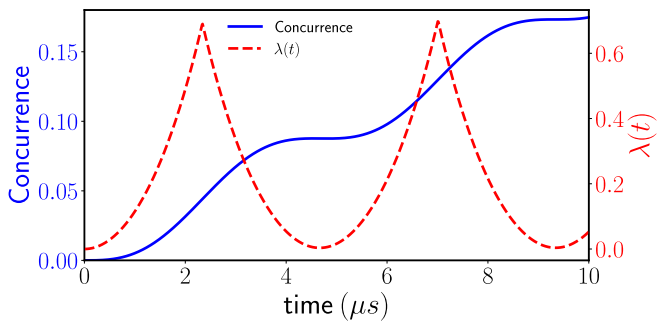}
\caption{Concurrence increases in the regions where the rate function $\lambda(t)$ becomes nonanalytic. $ B_{x} = 100 $ G and $B_{z} = 5 $ G}\label{Figure8}
\end{figure}

In Fig. \ref{Figure8} the Concurrence evidences the generation of quantum correlations between the two spins when the rate function presents nonanalytical points, in agreement with Ref.~\cite{jurcevic2017}.

Second, we pay attention to the generation of multipartite entanglement. To this end, we consider three nuclear spins. Then, the Hamiltonian for this new system, conditioned to the \NVm electron spin in the $m_{s} = 0$ manifold reads,
\begin{align}\label{H3S}
\hat{\mathcal{H}}&=\gamma_eB_{z}\sum_{i=1}^{3}\hat{I}_{z}^{(i)}+\gamma_nB_{x}\sum_{i=1}^{3}\hat{I}_{x}^{(i)}\nonumber\\
&+\frac{\beta_{12}}{2}\left((\hat{I}_{+}^{(1)}\hat{I}_{-}^{(2)}+\hat{I}_{-}^{(1)}\hat{I}_{+}^{(2)})-4\hat{I}_{z}^{(1)}\hat{I}_{z}^{(2)}\right)\nonumber\\
&+\frac{\beta_{23}}{2}\left((\hat{I}_{+}^{(2)}\hat{I}_{-}^{(3)}+\hat{I}_{-}^{(3)}\hat{I}_{+}^{(3)})-4\hat{I}_{z}^{(2)}\hat{I}_{z}^{(3)}\right).
\end{align}

For simplicity, we consider the spins in a 1D array configuration with first neighbors interaction and $\beta_{12} = \beta_{23}$. To quantify multipartite entanglement we use the Tangle ($\tau$)~\cite{Coffman2000},
\begin{equation}
\tau_{123}=\mathcal{C}^2_{1(23)}-\mathcal{C}_{12}^{2}-\mathcal{C}_{13}^{2},
\end{equation}

where $ \tau_{123} $ represents a residual entanglement of the collective three spins system~\cite{Coffman2000}, and $ \mathcal{C}_{1(23)}^2 = 2(1-\text{Tr}[\rho_{1}^{2}]) $ represents the entanglement between $^{13}\mbox{C}_1$ and pair $^{13}\mbox{C}_2- ^{13}\mbox{C}_3$. $\mathcal{C}_{12} $ and $ \mathcal{C}_{13}$ stand for the Concurrence of the bipartite systems given by Eq.~\eqref{Conc}.
\begin{figure}[h]
\centering
\includegraphics[width=230pt]{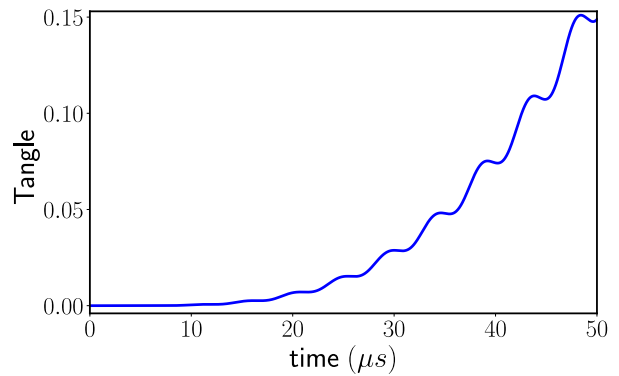}
\caption{Multipartite entanglement increases during the time evolution. After several critical points the plateaus show up, indicating a similar behavior with the Concurrence in Fig.~\ref{Figure8}. $ B_{x} = 100 $ G and $B_{z} = 5 $ G.}\label{Figure9}
\end{figure}

Fig.~\ref{Figure9} shows the evolution of the Tangle ($\tau_ {123}$) for the tripartite system made up of three \carb~nuclear spins. We note that at early evolution, the Tangle is approximately zero since it takes time for bipartite entanglement to built up first. After $\sim 15 $ $\mu$s, the Tangle increases, and the plateaus appear, coinciding with the minimum of the rate function $ \lambda (t)$ (similar to the Concurrence in Fig.~\ref{Figure8}). On the other hand, the critical points show an increasing Tangle. This implies that every time there is a critical point, a residual entanglement of the tripartite system is generated. 

\section{Scalability.}\label{Appendix_Scalability}

A key element for quantum phase transitions in a finite quantum system is the scalability with just a few spins. For our particular problem, the critical time in the DQPT must be universal regardless of the number of spins $N$. This assumption have been made when introducing rate function, $\mathcal{L}(t)=\exp(-N\Lambda(t))$, where $\Lambda(t)$ is independent of $N$. In this section, we analyze DQPT for different number of spins and show that DQPT holds for a fixed critical time.

\begin{figure}[h]
\centering
\includegraphics[width=240pt]{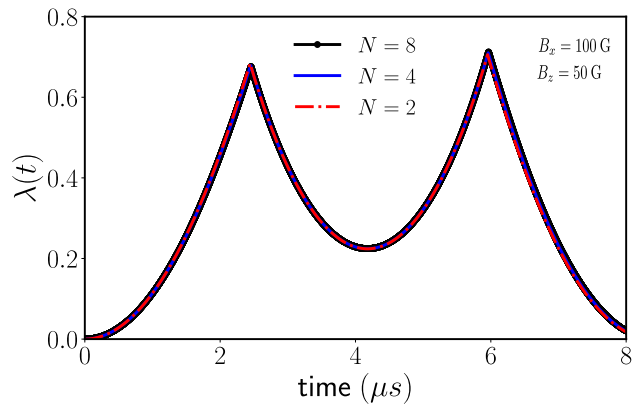}
\caption{Evolution of the rate function for different numbers of spins. The universality of the critical time holds.}\label{Figure10}
\end{figure}

In Fig.~\ref{Figure10} we show the evolution of the rate function for $N = \lbrace2, 4, 8\rbrace$ nuclear spins. Note that when increasing the number of spins, the nonanalytical points in the rate function show up around the same critical time as for the case of $N = 2$. For the calculations, we consider a fixed topology of a 1D array of nuclear spins with the same coupling strength between them.

\section{Fisher information for $\beta_{12}$ and $B_{x}=0$.}\label{Appendix_Fisher}

To begin with, we consider the Hamiltonian in the $m_s=0$ manifold given in Eq.~\eqref{eq8} for $B_x=0$,
\begin{align}
\hat{H}&=\gamma_nB_{z}(\hat{I}_{z}^{(1)}+\hat{I}_{z}^{(2)})\nonumber\\
&\quad+\frac{\beta_{12}}{2}\left((\hat{I}_{+}^{(1)}\hat{I}_{-}^{(2)}+\hat{I}_{-}^{(1)}\hat{I}_{+}^{(2)})-4\hat{I}_{z}^{(1)}\hat{I}_{z}^{(2)}\right).
\end{align}

The corresponding eigenstates and eigenvalues are,
\begin{align}
\ket{\psi_1}&=\ket{0\downarrow\downarrow},\quad E_{1}=-\left(\gamma_nBz+\frac{\beta_{12}}{2}\right)\nonumber\\
\ket{\psi_2}&=\ket{0\uparrow\uparrow},\quad E_{2}=\gamma_nBz-\frac{\beta_{12}}{2}\nonumber\\
\ket{\psi_3}&=\frac{1}{\sqrt{2}}\left(\ket{0\downarrow\uparrow}-\ket{0\uparrow\downarrow}\right),\quad E_{3}=0\nonumber\\
\ket{\psi_4}&=\frac{1}{\sqrt{2}}\left(\ket{0\downarrow\uparrow}+\ket{0\uparrow\downarrow}\right),\quad E_{3}=\beta_{12}
\end{align}

The initial state (probe state that increases the FI) is $\ket{0\uparrow\downarrow}$, that in the eigenstate basis can be written as
\begin{equation}
\ket{\psi(0)}= \ket{0\uparrow\downarrow}=\frac{1}{\sqrt{2}}\left(\ket{\psi_4}-\ket{\psi_3}\right).
\end{equation}

The evolution of this state reads,
\begin{align}
\ket{\psi(t)}&=e^{-i\hat{H}(\beta_{12})t}\ket{\psi(0)}
=\frac{1}{\sqrt{2}}\left(e^{-i\beta_{12}t}\ket{\psi_4}-\ket{\psi_3}\right).%\nonumber\\
%&=\frac{1}{2}\left(\left[e^{-i\beta_{12}t}-1\right]\ket{\downarrow\uparrow}+\left[e^{-i\beta_{12}t}+1\right]\ket{\uparrow\downarrow}\right)
\end{align}

In the bare basis the density matrix reads,
\begin{align}
\rho(t)&=\ket{\psi(t)}\bra{\psi(t)}\nonumber\\
&=\frac{1}{4}\left[\left(e^{-i\beta_{12}t}-1\right)\left(e^{i\beta_{12}t}-1\right)\ket{\downarrow\uparrow}\bra{\downarrow\uparrow}\right.\nonumber\\
&\quad\left.+\left(e^{-i\beta_{12}t}-1\right)\left(e^{i\beta_{12}t}+1\right)\ket{\downarrow\uparrow}\bra{\uparrow\downarrow}\right.\nonumber\\
&\quad\left.+\left(e^{-i\beta_{12}t}+1\right)\left(e^{i\beta_{12}t}-1\right)\ket{\uparrow\downarrow}\bra{\downarrow\uparrow}\right.\nonumber\\
&\quad\left.+\left(e^{-i\beta_{12}t}+1\right)\left(e^{i\beta_{12}t}+1\right)\ket{\uparrow\downarrow}\bra{\uparrow\downarrow}\right].
\end{align}

To obtain the reduced density matrix for $^{13}\mbox{C}_1$ we trace over the electron and $^{13}\mbox{C}_2$ degrees of freedom.
\begin{align}
\rho_{n_1}(t)&=\text{Tr}_{n_2}[\rho(t)]\nonumber\\
&=\bra{\uparrow}_{n_2}\rho(t)\ket{\uparrow}_{n_2}+\bra{\downarrow}_{n_2}\rho(t)\ket{\downarrow}_{n_2}\nonumber\\
%&=\frac{1}{4}\left(4\sin^{2}(\beta_{12}t/2)\ket{\downarrow}\bra{\downarrow}+4\cos^{2}(\beta_{12}t/2)\ket{\uparrow}\ket{\uparrow}\right)\nonumber\\
&=\sin^{2}\left(\frac{\beta_{12}t}{2}\right)\ket{\downarrow}\bra{\downarrow}+\cos^{2}\left(\frac{\beta_{12}t}{2}\right)\ket{\uparrow}\bra{\uparrow}.
% &=\frac{1}{4}\left[\underbrace{\left(e^{-i\beta_{12}t}-1\right)\left(e^{i\beta_{12}t}-1\right)}_{4\sin^{2}(\beta_{12}t/2)}\ket{\downarrow}\bra{\downarrow}\right.\nonumber\\
% &\quad\left.+\underbrace{\left(e^{-i\beta_{12}t}+1\right)\left(e^{i\beta_{12}t}+1\right)}_{4\cos^{2}(\beta_{12}t/2)}\ket{\uparrow}\ket{\uparrow}\right]\nonumber\\
% \rho_{n_1}(t)&=\sin^{2}\left(\frac{\beta_{12}t}{2}\right)\ket{\downarrow}\bra{\downarrow}+\cos^{2}\left(\frac{\beta_{12}t}{2}\right)\ket{\uparrow}\bra{\uparrow}
\end{align}

Next, we calculate the probability to find the nuclear spin in states $\ket{\uparrow}$ ($P_{\ket{\uparrow}}(\beta_{12}))$ and $\ket{\downarrow}$($P_{\ket{\downarrow}}(\beta_{12})$),
\begin{align}
P_{\ket{\uparrow}}(\beta_{12})&=\cos^{2}\left(\frac{\beta_{12}t}{2}\right),\\
P_{\ket{\downarrow}}(\beta_{12})&= 1-\cos^{2}\left(\frac{\beta_{12}t}{2}\right)=\sin^{2}\left(\frac{\beta_{12}t}{2}\right).
\end{align}

Finally, we replace the above expressions into the Fisher Information in Eq.~\eqref{FI2} and obtain
%\begin{equation}
% \frac{\partial P_{\ket{\uparrow}}(t,\beta_{12})}{\partial \beta_{12}}=-t\cos\left(\frac{\beta_{12}t}{2}\right)\sin\left(\frac{\beta_{12}t}{2}\right)
%\end{equation}
%
%Replacing the results obtained previously we obtain
%
% \begin{align}
% F(t,\beta_{12})= \frac{1}{\cos^{2}\left(\frac{\beta_{12}t}{2}\right)\sin^{2}\left(\frac{\beta_{12}t}{2}\right)}\left[-t\cos\left(\frac{\beta_{12}t}{2}\right)\sin\left(\frac{\beta_{12}t}{2}\right)\right]^{2}
% \end{align}
%
% Simplifying we obtain
\begin{equation}
\mbox{FI}(\beta_{12})=t^{2}.
\end{equation}

%Bibliografía
%\bibliographystyle{apsrev4-1}
%\bibliographystyle{plain}
%\bibliography{Referencias}

\end{document}